\documentclass{article}

\usepackage{graphicx}
\usepackage{epsfig}
\begin{document}

\title{\bf Higher-order geodesic deviations and orbital precession in a Kerr-Newman space-time}
\author{{Mohaddese Heydari-Fard$^{1}$ \thanks{Electronic address: m\_heydarifard@sbu.ac.ir}, Malihe Heydari-Fard$^{2}$\thanks{Electronic address: heydarifard@qom.ac.ir} and Hamid Reza Sepangi$^{1}$\thanks{Electronic address: hr-sepangi@sbu.ac.ir}}\\ {\small \emph{$^{1}$ Department of Physics, Shahid Beheshti University, G. C., Evin, Tehran 19839, Iran}}
\\{\small \emph{$^{2}$ Department of Physics, The University of Qom, 3716146611, Qom, Iran}}}

\maketitle

\begin{abstract}
A novel approximation method in studying the perihelion precession and planetary orbits in general relativity is to use geodesic deviation equations of first and high-orders, proposed by Kerner et.al. Using higher-order geodesic deviation approach, we generalize the calculation of orbital precession and the elliptical trajectory of neutral test particles to Kerr$-$Newman space-times. One of the advantage of this method is that, for small eccentricities, one obtains trajectories of planets without using Newtonian and post-Newtonian approximations for arbitrary values of quantity ${G M}/{R c^2}$.
\vspace{5mm}
\vspace{1mm}
\end{abstract}

\section{Introduction}
Over the past century, the theory of General Relativity (GR) has paved the way for many predictions including, planet's perihelion advance \cite{1}, bending of light when passing close by the sun, gravitational red shift \cite{2}, accelerating expansion of the universe \cite{3} and the existence of gravitational waves \cite{4}, which have been confirmed by observational evidence \cite{5}.

The precession of planet's perihelion is a famous problem in general relativity, see Figure 1. The residual perihelion advance of Mercury is 43 arc-seconds per century. Newtonian mechanics accounts for 531 arc-seconds per century by computing the perturbing influence of other planets in the solar system, while the predicted value of precession by observation is 574 arc-second per century. This residual value was obtained from observation by Urbain Le Verrier in 1859. To explain this discrepancy Li Verrier suggested a new planet (named Vulcon) near the sun which turned out not to exist \cite{Straumann}.

Einstein \cite{1} was the first to investigate the motion of planets as test particles moving along geodesics in a Schwarzschild space-time which, as is well known, leads to elliptic integrals. By expanding the integrand in powers of the small parameter $\frac{GM}{r}$ and keeping only linear terms in the expansion, he found the well-known formula for perihelion advance \cite{Weinberg}
\begin{equation}
\Delta{\varphi}=\frac{6\pi GM}{a(1-e^{2})},\label{0}
\end{equation}
where $G$ is the gravitational constant, $M$ is the mass of the central object, $a$ and $e$ are the major-semi axis and the eccentricity of the ellipse respectively. For small eccentricities the above relation can be expanded as a power series
\begin{equation}
\Delta{\varphi}=\frac{6\pi GM}{a} (1+e^{2}+e^{4}+e^{6}+\cdots).\label{1}
\end{equation}
Interestingly, even for planet Mercury with greatest orbital eccentricity, the series converges rapidly such that the inclusion of the second term, that is $(1+e^2)$, will lead to a result that differs only by 0.18\% from the result predicted by relation (\ref{0}), which is below the actual error bars. This would suggest that if orbital trajectories of test particles can be presented in a power series of eccentricity, rapid conversion and precise predictions will be achieved. This is what will be done in what follows, using the geodesic deviation method.

A novel approximation method for calculating the perihelion precession and finding trajectory of test particles in various gravitational fields without introducing the Newtonian and Post-Newtonian approximation has been suggested by Kerner {\it{et.al.}} \cite{Kerner}. In two previous papers, the authors calculate the perihelion precession and trajectory of test particles in Schwarzschild and Kerr space-times \cite{Kerner2}--\cite{Balakin}. Our aim is to generalize the geodesic deviation method to the Kerr-Newmann geometry. The motion of neutral and charged test particles along closed orbits in Kerr-Newman metric has been analysed in an exhaustive manner in many papers \cite{KerrNewman}--\cite{KerrNewman8}. For further insight on the alternative methods in calculating the orbital precession and planetary orbits in Schwarzschild, Reissner-Nordstrom, Kerr and Kerr-Newman space-times see \cite{Method}--\cite{Method7}.

In this paper, instead of using Einstein's direct method and writing the integrand in term of the small parameter $\frac{GM}{r}$ or using Newtonian and Post-Newtonain approximation, we consider a particulary simple geodesic, namely the circular orbit. We then add the first, second and higher-order deviations to such an orbit, thereby  obtaining the exact trajectory of the neutral test particles. The advantage of this approach is in finding explicit form of the trajectory for arbitrary values of parameter $\frac{GM}{r}$ i.e., in the vicinity of very massive and compact bodies, and also to take into account the full relativistic effects without perturbing the curvature of the space-time. However, this method is valid for small eccentricities. In addition to the method presented here, generalization of geodesic deviation equations to second and third orders are  presented in \cite{new1}-\cite{new3}. Also in \cite{new4}, a derivation based on covariant bitensor formalism \cite{new5}-\cite{new7} for higher orders geodesic deviation has been presented.
\begin{figure}
\centering
\includegraphics[width=3.0in]{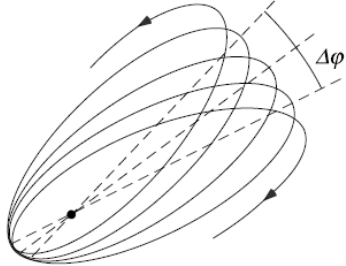}
\caption{Planetary orbit in GR. Figure from \cite{Straumann}.}
\label{Planetary}
\end{figure}

The paper is arranged as follows. In section 2 we give a brief review of the method used here. The circular motion in Kerr-Newman metric is explained in section 3. In section 4, we study the first geodesic deviation around this circular orbit. To obtain the trajectory of test particles in Kerr-Newman field,  we derive the second and third-order geodesic deviations in section 5. Finally, conclusions are presented in section 6.

\section{Geodesic deviation method}
One of the most important equations in General Relativity is the geodesic deviation equation which, for a neutral particle is given by \cite{Weinberg,Misner}
\begin{equation}
\frac{D^{2}n^{\mu}}{D\tau ^{2}}=R^{\mu}_{\,\,\,\,
\lambda\nu\kappa}u^{\lambda}n^{\nu}u^{\kappa}.\label{2}
\end{equation}
As is shown in the Figure 2, $u^{\alpha}(\tau, \lambda)=\frac{\partial x^{\alpha}}{\partial \tau}$ is a time-like tangent vector along the geodesic and $n^{\alpha}(\tau, \lambda)=\frac{\partial x^{\alpha}}{\partial \lambda}$ is the deviation four-vector. Also, $\frac{D}{D\tau}$ is the covariant derivative along the proper time $\tau$ and $R^{\mu}_{\,\,\,\,\lambda\nu\kappa}$ is the Riemann tensor of the space-time. It is often convenient to rewrite the above equation in the non-trivial covariant form
\begin{equation}
\frac{d^{2}n^{\mu}}{d\tau ^{2}}+2\Gamma ^{\mu}_{\,\,\,\,\kappa\nu}u^{\kappa}\frac{dn^{\nu}}{d\tau}+\partial _{\nu}\Gamma ^{\mu}_{\,\,\,\,\kappa\sigma}u^{\kappa}u^{\sigma}n^{\nu}=0.\label{3}
\end{equation}
Now, we use the geodesic deviation method to construct one geodesic from an initial one by expansion in a series of powers of small parameter \cite{Kerner}
\begin{equation}
x^{\mu}(\tau, \lambda)=x^{\mu}(\tau, \lambda_{0})+(\lambda-\lambda_{0})\frac{\partial x^{\mu}}{\partial \lambda}|_{(\tau, \lambda_{0})}+\frac{1}{2!}(\lambda-\lambda_{0})^{2}\frac{\partial ^{2} x^{\mu}}{\partial \lambda^{2}}|_{(\tau, \lambda_{0})}+\cdots. \label{4}
\end{equation}
Our aim is to obtain an expression in terms of deviation vectors. In the above equation, the second term, $\frac{\partial x^{\mu}}{\partial \lambda}$, is the definition of deviation vector and shows deviation from first-order geodesic. However, in third term, $\frac{\partial ^{2} x^{\mu}}{\partial \lambda^{2}}$, is not a vector anymore. One is therefore required to change this term to an expression showing the second-order geodesic deviation. To this end, we define the vector $b^{\mu}$
\begin{equation}
b^{\mu}=\frac{Dn^{\mu}}{D\lambda}=\frac{\partial n^{\mu}}{\partial \lambda}+\Gamma ^{\mu}_{\lambda\nu}n^{\lambda}n^{\nu}.\label{5}
\end{equation}
Now, by substituting this in equation (\ref{4}), one obtains an expression in terms of the order of vector deviations
\begin{equation}
x^{\mu}(\tau, \lambda)=x^{\mu}(\tau, \lambda_{0})+\epsilon n^{\mu}+\frac{1}{2!}\epsilon ^{2}(b^{\mu}-\Gamma ^{\mu}_{\lambda\nu}n^{\lambda}n^{\nu})+\cdots ,\label{6}
\end{equation}
where $\epsilon=(\lambda-\lambda_0)$. For simplicity we introduce $\delta ^{n} x^{\mu}(\tau)$ as the $n$-th order of deviation, then equation (\ref{6})  takes the form
\begin{equation}
x^{\mu}(\tau , \lambda)=x^{\mu}_{0}(\tau)+\delta x^{\mu}(\tau)+\frac{1}{2}\delta ^{2}x^{\mu}(\tau)+\cdots ,\label{7}
\end{equation}
where $\delta x^{\mu}(\tau)=\epsilon n_{0}^{\mu}(s)$ and $\delta ^{2}x^{\mu}(\tau)=\epsilon^{2}(b^{\mu}-\Gamma^{\mu}_{\nu\lambda}n^{\nu}n^{\lambda})$ are first and second-order geodesic deviations, respectively. With straightforward calculations, one finds that vector $ b^{\mu} $ satisfies
 \begin{eqnarray}
\frac{D^{2}b^{\mu}}{Ds^{2}}+R^{\mu}_{\,\,\rho\lambda\sigma}u^{\lambda}b^{\rho}u^{\sigma} = (R^{\mu}_{\,\,\rho\lambda\sigma ;\nu}-R^{\mu}_{\,\,\sigma\nu\rho ;\lambda})u^{\lambda}u^{\sigma}n^{\rho}n^{\nu}+4R^{\mu}_{\,\,\rho\lambda\sigma}u^{\lambda}\frac{Dn^{\rho}}{Ds}n^{\sigma}.\label{8}
\end{eqnarray}
Rewriting the above equation in a non-trivial covariant form, we have
\begin{eqnarray}
\frac{d^{2}b^{\mu}}{d\tau ^{2}}&+&2\Gamma ^{\mu}_{\,\,\,\,\kappa\nu}u^{\kappa}\frac{db^{\nu}}{d\tau}+\partial _{\nu}\Gamma ^{\mu}_{\,\,\,\,\kappa\sigma}u^{\kappa}u^{\sigma}b^{\nu}=4(\partial _{\lambda}\Gamma ^{\mu}_{\,\,\,\,\sigma\rho}+\Gamma ^{\nu}_{\,\,\,\,\sigma\rho}\Gamma ^{\mu}_{\,\,\,\,\lambda\nu})\frac{dn^{\sigma}}{d\tau}(u^{\lambda}n^{\rho}-u^{\rho}n^{\lambda})\nonumber\\
&+&(\Gamma ^{\tau}_{\,\,\,\,\sigma\nu}\partial _{\tau}\Gamma ^{\mu}_{\,\,\,\,\lambda\rho}+2\Gamma ^{\mu}_{\,\,\,\,\lambda\tau}\partial _{\rho}\Gamma ^{\tau}_{\,\,\,\,\sigma\nu}-\partial _{\nu}\partial _{\sigma}\Gamma ^{\mu}_{\,\,\,\,\lambda\rho})(u^{\lambda}u^{\rho}n^{\sigma}n^{\nu}-u^{\sigma}u^{\nu}n^{\lambda}n^{\rho}),\label{9}
\end{eqnarray}
which is called the second-order geodesic deviation equation. Generalization of this equation for the case of the charged particles has recently  been obtained in \cite{Heydari-Fard}. It is worth noting that the left-hand sides of equations (\ref{3}) and (\ref{9}) are the same, but in the latter there are some extra terms. It can easily be shown that by extending the method to third and higher orders, this general pattern persists. In other words, the geodesic deviation equations of higher order have the same left-hand side and differ from each other in the extra terms appearing on the right-hand side which depend on the solutions of lower orders.
\begin{figure}
\centering
\includegraphics[width=4.0in]{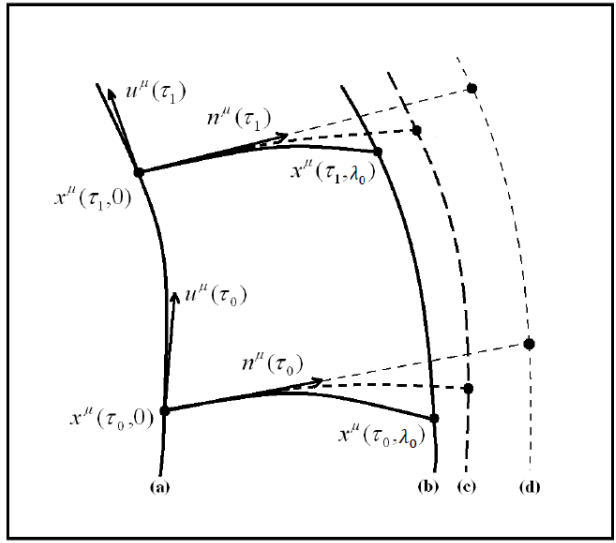}
\caption{ Deviation of two nearby geodesics in a gravitational field. Lines (a)
and (b) represent the central geodesic $\lambda = 0$ and the nearby geodesic
$\lambda = \lambda _0$, respectively; lines (c) and (d) show the corresponding first
and second-order approximations to the nearby geodesic (b). Also, $u^{\mu}$ is the unit tangent vector to the central world-line and $n^{\mu}$ is the separation vector to the curve $\tau  = \mbox{const}$. Figure from \cite{Baskaran}.}
\label{Deviation}
\end{figure}

In the next sections, by considering a circular orbit of radius $ R $ as the original geodesic $ x^{\mu}_{0}(\tau ,\lambda) $ for a neutral particle in Kerr-Newman space-time and using equations (\ref{3}) and (\ref{9}), we obtain the components of deviation vectors $ n^{\mu} $ and $ b^{\mu} $ in such space-times.

\section{Circular orbits in Kerr-Newman metric}
Let us now study the circular orbits outside a massive, charged, rotating source \cite{Islam}. This space-time which is described by the Kerr-Newman metric is an exact, axis-symmetric and asymptotically flat solution of the Einstein-Maxwell equations. In the Boyer$-$Lindquist coordinates (with $c=G=1$) the line-element is given by
\begin{eqnarray}
ds^{2}&=&-\frac{(\Delta   -a^{2}\sin^{2}\theta)}{\rho^{2}}dt^{2}+\frac{\rho^{2}}{\Delta}dr^{2}+\rho^{2}d\theta ^{2}
\nonumber\\&+&2\frac{a \sin^{2}\theta}{\rho^{2}}(\Delta -r^{2}-a^
{2})dt d\varphi -\frac{\sin^{2}\theta}{\rho^{2}}\left( \Delta a^{2}\sin^{2}\theta - (r^{2}+a^{2})^{2}\right) d\varphi ^{2},\label{10}
\end{eqnarray}
with
\begin{eqnarray}
\rho^2 = r^{2}+a^{2}\cos^{2}\theta ,\hspace{.5cm} \Delta = r^{2}+a^{2}+Q^{2}-2Mr,\label{11}
\end{eqnarray}
where parameters $M$, $a$ and $Q$ stand for mass, angular momentum per unit mass and charge of the black-hole. As is clear, in the limiting cases $Q=0$ and $a=0$ the Kerr-Newman metric reduces to the Kerr and Reissner-Nordstrom metrics. Also it reduces to the Schwarzschild metric for $a=0$, $Q=0$. For further insight, see \cite{Misner}.

Since the Kerr-Newman metric has no explicit dependence on $t$ and $\varphi$ coordinates, two conserved quantities, namely energy ($\varepsilon$) and angular momentum ($l$) can be obtained from Euler-Lagrange equations
\begin{eqnarray}
u^t=\dot{t}=\varepsilon -\frac{(Q^2-2Mr)[(r^2+a^2)\varepsilon -al]}{\Delta\rho ^2},\label{12}
\end{eqnarray}
\begin{eqnarray}
u^{\varphi}=\dot{\varphi}=\frac{l}{\rho ^2\sin ^2\theta} -\frac{a[(Q^2-2Mr)\varepsilon +al]}{\Delta\rho ^2},\label{13}
\end{eqnarray}
where a dot represents derivative with respect to the proper time $\tau$. For the radial coordinate $r$ we find
\begin{eqnarray}
\rho ^2\dot{r}^2=[(r^2+a^2)\varepsilon -al]^2-\Delta [r^2+(l-a\varepsilon)^2],\label{14}
\end{eqnarray}
from which we obtain a relation for the radial acceleration
\begin{eqnarray}
\rho ^2\ddot{r}=2r\varepsilon[(r^2+a^2)\varepsilon -al]-r\Delta -(r-M)[r^2+(l-a\varepsilon)^2].
\label{15}
\end{eqnarray}
Now, let us consider a neutral test particle with mass $m\ll M$ along a circular orbit of radius $R$ in the equatorial plane. Then using constraints $\theta=\frac{\pi}{2}$ and $ \dot{r}=\ddot{r}=0 $ in the above equations, components of the four-velocity vector are given by
\begin{eqnarray}
u^{r}&=&0, \hspace{.5cm} u^{\theta}=0,\nonumber\\
u^{t}&=&\frac{1+\frac{a}{R}\sqrt{\frac{M}{R}-\frac{Q^{2}}{R^{2}}}}{\sqrt{1-\frac{3M}{R}+2\frac{Q^{2}}{R^{2}}+
\frac{2a}{R}\sqrt{\frac{M}{R}-\frac{Q^{2}}{R^{2}}}}},
\nonumber\\
u^{\varphi}&=&\omega _0=\frac{\sqrt{\frac{M}{R}-\frac{Q^{2}}{R^{2}}}}{R\sqrt{1-\frac{3M}{R}+2\frac{Q^{2}}{R^{2}}+
\frac{2a}{R}\sqrt{\frac{M}{R}-\frac{Q^{2}}{R^{2}}}}}.\label{16}
\end{eqnarray}
It is clear that with increasing charge and rotation parameters the angular frequency of the circular orbit (unperturbed initial orbit), $ \omega _0 $, decreases.

\section{The first-order geodesic deviation around circular orbits}
Now, we are ready to solve the first-order geodesic deviations in a Kerr-Newman space-time. To this end, in equation (\ref{3}) the radial coordinate $r$ is set to constant $R$, the four-velocities are given by equation (\ref{16}), and our calculations are restricted to the equatorial plane, $\theta=\frac{\pi}{2}$. Thus the first-order geodesic deviation equation takes the form
\begin{equation}
\left( \begin{array}{cccc} \frac{d^2}{d\tau^2} & \alpha\frac{d}{d\tau} & 0 & 0  \\  \beta\frac{d}{d\tau} & \frac{d^2}{d\tau^2}+\kappa & 0 &\gamma\frac{d}{d\tau}\\  0 & 0 & \frac{d^2}{d\tau^2}+\delta & 0
\\ 0 & \eta\frac{d}{d\tau} & 0 & \frac{d^2}{d\tau^2} \end{array} \right)\left( \begin{array}{cc}n^{t}\\ n^{r}\\n^{\theta} \\ n^{\varphi}\end{array} \right)
=\left( \begin{array}{cc}0\\0\\ 0 \\0\end{array}\right),\label{17}
\end{equation}
where
\begin{eqnarray}
&\alpha = \frac{2(MR-Q^2)f_5}{R^3f_2\sqrt{f_1}},\hspace{.5cm} \beta = \frac{2(MR-Q^2)f_2}{R^3\sqrt{f_1}},\hspace{.5cm} \kappa = -\frac{3(MR-\frac{4}{3}Q^2)f_2}{R^4f_1},
\nonumber\\
&\gamma =-\frac{2\sqrt{MR-Q^2}}{R}(1+\frac{a\sqrt{MR-Q^2}}{R^2})\frac{f2}{\sqrt{f1}},\hspace{.5cm} \delta =\frac{(MR-Q^2)f3}{R^4f1},\hspace{.5cm} \eta =\frac{2\sqrt{MR-Q^2}f4}{R^3f2\sqrt{f1}},\label{18}
\end{eqnarray}
with
\begin{eqnarray}
f_1 &=&1- \frac{3M}{R}+\frac{2a}{R}\sqrt{\frac{M}{R}-\frac{Q^2}{R^2}}+\frac{2Q^2}{R^2},\hspace{.5cm} f_2 =1- \frac{2M}{R}+\frac{a^2}{R^2}+\frac{Q^2}{R^2},\nonumber\\
f_3 &=&1-\frac{4a}{R}\sqrt{\frac{M}{R}-\frac{Q^2}{R^2}}+\frac{3a^2}{R^2}+
\frac{a^2Q^2}{R^4}\left(\frac{M}{R}-\frac{Q^2}{R^2}\right)^{-1},\hspace{.5cm} f_4=1-\frac{2M}{R}+\frac{a}{R}\sqrt{\frac{M}{R}-\frac{Q^2}{R^2}}+\frac{Q^2}{R^2},\nonumber\\
f_5 &=&1 -\frac{2a}{R}\sqrt{\frac{M}{R}-\frac{Q^2}{R^2}}+\frac{a^2}{R^2}-
\frac{aQ^2}{R^3}\left(\frac{M}{R}-\frac{Q^2}{R^2}\right)^{-\frac{1}{2}}.\label{19}
\end{eqnarray}
From equation (\ref{17}), we can see that $n^{\theta}$ component of geodesic deviation equation is independent of variables $ n^{t} $,$ n^{r} $ and $ n^{\varphi} $. The harmonic oscillator equation for $ n^{\theta} $ has an angular frequency given by
\begin{eqnarray}
\omega_{\theta}=\frac{\sqrt{\frac{M}{R}-\frac{Q^2}{R^2}}}{R}\sqrt{\frac{f_3}{f_1}},\label{20}
\end{eqnarray}
with the following solution
\begin{equation}
n^{\theta} =-n_0^{\theta}\cos(\omega_{\theta}\tau).\label{21}
\end{equation}
In the Schwarzschild and the Kerr limit, the above result reduces to that of references \cite{Kerner} and \cite{Colistete}, respectively. It is worth noting that in the Schwarzschild limit we have $ \omega_{\theta}=\omega_{0} $, namely the period is equal to the period of the planetary motion which can be interpreted as a pure coordinate effect so in this case we can take $ n^{\theta}=0 $.

Now, by eliminating derivatives of $ n^{\varphi} $ and $ n^{t} $ in differential equation for $ n^r $, we find the following harmonic oscillator equation
\begin{equation}
\frac{d^2n^r}{ds^2}+\omega^{2}n^{r}=0,\label{22}
\end{equation}
with the characteristic frequency
\begin{equation}
\omega =\frac{\sqrt{\frac{M}{R}-\frac{Q^2}{R^2}}}{R}\sqrt{\frac{f_6}{f_1}},\label{23}
\end{equation}
where
\begin{equation}
f_6=1-\frac{6M}{R}+\frac{8a}{R}\sqrt{\frac{M}{R}-\frac{Q^2}{R^2}}-\frac{3a^2}{R^2}+
\frac{4Q^2}{R^2}+\frac{Q^2}{R^2}\left(1-\frac{M}{R}+\frac{a^2}{R^2}\right)\left(\frac{M}{R}-
\frac{Q^2}{R^2}\right)^{-1}.\label{24}
\end{equation}
\begin{equation}
n^r=-n_0^r\cos(\omega \tau),\label{25}
\end{equation}
resulting in perihelion and aphelion corresponding to $\tau=0$ and $\tau=\frac{\pi}{\omega}$, respectively. We note that the discrepancy between the two angular frequencies $\omega_0$ and $\omega$ is what causes the perihelion precession. The solutions $n^{\varphi}$ and $n^t $ are given
by
\begin{eqnarray}
n^{\varphi}&=&n_0^{\varphi}\sin(\omega \tau),\nonumber\\
n^t&=&n_0^t\sin(\omega \tau),\label{26}
\end{eqnarray}
where the amplitudes depend on $n_0^r$
\begin{eqnarray}
n_{0}^{\varphi}&=&\frac{2n_{0}^{r}}{R}\frac{f_{4}}{f_{2}\sqrt{f_{6}}},\nonumber\\
n_{0}^{t}&=&2n^{r}_{0}\sqrt{\frac{M}{R}-\frac{Q^{2}}{R^{2}}}\frac{f_{5}}{f_{2}\sqrt{f_{6}}}.\label{27}
\end{eqnarray}

Now, these first-order deviations (linear approximation) should be added to the unperturbed circular orbit (zeroth-order approximation) to obtain the new trajectory as follows
\begin{eqnarray}
r = R-n_0^r\cos(\omega \tau)=R[1-\frac{n_0^r}{R}\cos(\omega \tau)],\label{28}
\end{eqnarray}
\begin{eqnarray}
\theta &=& \frac{\pi}{2}-n_0^{\theta}\cos(\omega_{\theta}
\tau),\label{29}
\end{eqnarray}
\begin{eqnarray}
\varphi &=& \omega _0\tau+n_0^{\varphi}\sin(\omega \tau)=\omega _0 \tau+\frac{2n_{0}^{r}}{R}\frac{f_{4}}{f_{2}\sqrt{f_{6}}}\sin(\omega \tau),
\end{eqnarray}\label{30}
\begin{eqnarray}
t &=& u^t \tau +n_0^t\sin(\omega \tau)=u^t \tau+2n^{r}_{0}\sqrt{\frac{M}{R}-\frac{Q^{2}}{R^{2}}}\frac{f_{5}}{f_{2}\sqrt{f_{6}}}\sin(\omega \tau).\label{31}
\end{eqnarray}
If we compare equation (\ref{28}) with the Kepler result \cite{Kepler}
\begin{eqnarray}
r(t)=\frac{a(1-e^2)}{1+e\cos(\omega_{0}t)}\simeq a\left[1-e\cos(\omega_0 t)\right],\label{32}
\end{eqnarray}
we may identify the eccentricity $ e $ with $ \frac{n_0^r}{R} $ and the greater half-axis $ a $ with $ R $. Also, we can obtain an approximation for the orbital precession of neutral test particles in the equatorial plane of the Kerr-Newman space-time, using first-order deviations
\begin{eqnarray}
\bigtriangleup\varphi &=&2\pi \left(\frac{\omega_0}{\omega}-1\right)=\left(\frac{6\pi M}{R}+\frac{27\pi M^{2}}{R^{2}}+\frac{135\pi M^3}{R^3}+\cdots\right)-\frac{Q^2}{R^2}\left(\frac{\pi R}{M}+12\pi +\frac{189\pi M}{2R}\right.\nonumber\\
&+&\left.\frac{675\pi M^2}{R^2}+\cdots\right)-\frac{a\sqrt{M}}{R^{\frac{3}{2}}}\left(8\pi +\frac{72\pi M}{R}+\frac{540\pi M^2}{R^2}+\cdots\right)+\frac{a^2}{R^2}\left(3\pi +\frac{75\pi M}{R}\right.\nonumber\\
&+&\left.\frac{1845\pi M^2}{R^2}+\cdots\right)+\frac{a\sqrt{M}Q^2}{R^{7/2}} \left(216\pi +\frac{2430\pi M}{R}+\frac{5670\pi M^2}{R^2}+\cdots\right)\nonumber\\
&-&\frac{a^2Q^2}{R^4}\left(258\pi +\frac{18315\pi M}{4R}+\frac{70875\pi M^2}{4R^2}+\cdots\right)+\cdots .\label{33}
\end{eqnarray}
As one might expect, for the most general black-hole, i.e. Kerr-Newman, in addition to the mass,   the precession of perihelion is also modified by the angular momentum and electric charge of the source. In the method presented here, the orbital precession contains higher-order terms, $\frac{M}{R}$,  which demonstrate higher accuracy as compared to that of  Einstein's direct method. In the lowest-order of approximation the orbital precession can thus be written as
 \begin{equation}
\bigtriangleup\varphi =\frac{6\pi M}{R}-\frac{8\pi a \sqrt{M}}{R^{3/2}}+\frac{3\pi a^2}{R^2}-\frac{\pi Q^2}{R M}-\frac{258 \pi a^2 Q^2}{R^4}+\frac{216 \pi a\sqrt{M} Q^2}{R^{7/2}}+\cdots .\label{34}
\end{equation}

It is now interesting to investigate how different parameters of a central body influence the perihelion advance. The first term is the standard value for the perihelion precession of a test particle in the Schwarzschild field. The second and third terms are the effect of the rotation of central body. If the rotation of central body and the orbital motion of test particle are in the same or opposite directions, the linear term in the rotation parameter $a$ can make the perihelion advance to increase or decrease, respectively. Also, the quadratic term in $a$ always leads to higher precession since such extra energy would add extra mass to the central body according to the equivalence principle. Moreover, an extra term for the Reissner-Nordstorm space-time is observed and since the charge parameter appears in the quadratic terms, the sign of $Q$  is not important and the charge of the black-hole causes the precession to diminish (the charge parameter of the black hole seems to induce repulsive effects \cite{RN}). As can be seen from the above equation, orbital precession contains a new term $\frac{216 \pi a\sqrt{M} Q^2}{R^{7/2}}$, namely the Kerr-Newman effect, which couples the gravitational mass of the source, angular momentum and charge and tends to enhance the perihelion advance.

In the absence of charge, equation (\ref{33}) reduces to the usual orbital precession in Kerr space-time \cite{Colistete}. Also if the rotation of the gravitational source vanishes and the source becomes neutral, the result reduces to that of the perihelion precession in a Schwarzschild field, that is
\begin{equation}
\bigtriangleup\varphi =\frac{6\pi M}{R}+\frac{27\pi M^{2}}{R^{2}}+\frac{135\pi M^{3}}{R^{3}}+\cdots ,\label{35}
\end{equation}
which corresponds to equation (48) of reference \cite{Kerner}. By comparing the above equation to equation (\ref{1}), we find extra quadratic and cubic terms of the form $\frac{M}{R}$ which would improve the accuracy. It also shows that the geodesic deviation method can be used in the vicinity of massive and compact objects.

It is also clear that to first-order deviations  we only keep terms up to $e^2$. In what follows we derive and solve higher-order geodesic deviations and obtain non-linear terms involving $e^2=(\frac{n_0^r}{R})^2$, together with other higher-order effects.

\section{Second and third-order geodesic deviation}
For calculating the trajectory of neutral test particles in a Kerr-Newman space-time, we need to obtain a set of deviation vectors $(n^{\mu}(\tau), b^{\mu}(\tau), \cdots)$ on the reference geodesic. To this end, we insert the complete solution of the first-order deviation vector, equations (\ref{21}), (\ref{25}), and (\ref{26}) into equation (\ref{9}), leading to a set of linear second-order differential equations for deviation vector $b^{\mu}(\tau)$
\begin{equation}
\left( \begin{array}{cccc} \frac{d^2}{d\tau^2} & \alpha\frac{d}{d\tau} & 0 & 0  \\  \beta\frac{d}{d\tau} & \frac{d^2}{d\tau^2}+\kappa & 0 &\gamma\frac{d}{d\tau}\\  0 & 0 & \frac{d^2}{d\tau^2}+\delta & 0
\\ 0 & \eta\frac{d}{d\tau} & 0 & \frac{d^2}{d\tau^2} \end{array} \right)
\left( \begin{array}{cc}b^{t}\\ b^{r}\\b^{\theta} \\ b^{\varphi}\end{array} \right)={\epsilon}^{2}\left( \begin{array}{cc}C^{t}\\ C^{r} \\C^{\theta}\\ C^{\varphi}\end{array} \right),\label{36}
\end{equation}
where the coefficient $ \epsilon $ shows dependence of $ \delta^2x^{\mu} $ on the amplitude $ n_0^r $ ($ n_0^{\theta}$), and constants $C^{r}$, $C^{\theta}$, $C^{\varphi}$ and $C^{t}$ contain quantities depending on $M$, $R$, $a$, $Q$ and $\omega$ , $\omega_{0}$ and periodic functions of the form $\sin(2\omega\tau)$, $\cos(2\omega\tau)$ $\sin(2\omega_{\theta}\tau)$ and $\cos(2\omega_{\theta}\tau)$
\begin{equation}
C^{t}=C^t_{r}\sin(\omega \tau)+C^t_{\theta} \sin(\omega_{\theta}\tau),\label{37}
\end{equation}
\begin{equation}
C^{r}=C^r_{0}+C^r_{r}\cos(\omega \tau)+C^r_{\theta} \cos(\omega_{\theta}\tau),\label{38}
\end{equation}
\begin{equation}
C^{\theta}=C^{\theta}_{r}\cos[(\omega-\omega_{\theta}) \tau]+C^{\theta}_{\theta} \cos[(\omega+\omega_{\theta})\tau],\label{39}
\end{equation}
\begin{equation}
C^{\varphi}=C^{\varphi}_{r}\sin(\omega \tau)+C^{\varphi}_ {\theta} \sin(\omega_{\theta}\tau).\label{40}
\end{equation}
The solution for $b^{\mu}(\tau)$ has two terms, the general oscillatory solution with angular frequency $\omega$ and $\omega_{\theta}$ that accounted for $ n^{\mu} $, and the second particular solution which includes the oscillatory terms with angular frequency $2\omega$, $2\omega_{\theta}$, $(\omega+\omega_{\theta})$ and $(\omega-\omega_{\theta})$ as follows
\begin{eqnarray}
b^{t}(\tau)&=&\frac{(n_0^{\theta})^2\sqrt{\frac{M}{R}-\frac{Q^2}{R^2}}}{\sqrt{f_3}}\sin(2\omega _{\theta}\tau)\left[\frac{a^2}{R}+\frac{a^2Q^2}{2R^3}\left(\frac{M}{R}-\frac{Q^2}{R^2}\right)^{-1}\right]\nonumber\\
&+&\frac{(n_0^{r})^2\sqrt{\frac{M}{R}-\frac{Q^2}{R^2}}}{Rf_2f_6^{\frac{3}{2}}}\sin(2\omega\tau)\bigg\lbrace\left[\left(2-\frac{13M}{R}+\frac{11Q^2}{R^2}\right)+\frac{1}{2}\left(\frac{3Q^2}{R^2}-\frac{Q^4}{R^4}\right)\left(\frac{M}{R}-\frac{Q^2}{R^2}\right)^{-1}\right.\nonumber\\
&-&\left.\frac{1}{2}\left(\frac{Q^4}{R^4}-\frac{Q^6}{R^6}\right)\left(\frac{M}{R}-\frac{Q^2}{R^2}\right)^{-2}\right]+\frac{a}{R}\sqrt{\frac{M}{R}-\frac{Q^2}{R^2}}\left[\left(17+\frac{8M}{R}-\frac{3Q^2}{R^2}\right)-\frac{1}{2}\left(\frac{15Q^2}{R^2}-\frac{3Q^4}{R^4}\right)\right.\nonumber\\
&&\left.\left(\frac{M}{R}-\frac{Q^2}{R^2}\right)^{-1}-\frac{1}{2}\left(\frac{Q^4}{R^4}-\frac{Q^6}{R^6}\right)\left(\frac{M}{R}-\frac{Q^2}{R^2}\right)^{-2}\right]-\frac{a^2}{R^2}\left[\left(5+\frac{25M}{R}-\frac{21Q^2}{R^2}\right)\right.\nonumber\\
&-&\left.\frac{1}{2}\left(\frac{12Q^2}{R^2}+\frac{3Q^4}{R^4}\right)\left(\frac{M}{R}-\frac{Q^2}{R^2}\right)^{-1}+\frac{1}{2}\left(\frac{2Q^4}{R^4}-\frac{Q^6}{R^6}\right)\left(\frac{M}{R}-\frac{Q^2}{R^2}\right)^{-2}\right]\nonumber\\
&+&\frac{a^3}{R^3}\sqrt{\frac{M}{R}-\frac{Q^2}{R^2}}\left[23-\frac{13Q^2}{2R^2}\left(\frac{M}{R}-\frac{Q^2}{R^2}\right)^{-1}-\frac{3Q^4}{2R^4}\left(\frac{M}{R}-\frac{Q^2}{R^2}\right)^{-2}\right]\nonumber\\
&-&\frac{a^4}{R^4}\left[7-\frac{9Q^2}{2R^2}\left(\frac{M}{R}-\frac{Q^2}{R^2}\right)^{-1}+\frac{Q^4}{2R^4}\left(\frac{M}{R}-\frac{Q^2}{R^2}\right)^{-2}\right]\bigg\rbrace
,\label{41}\end{eqnarray}
\begin{eqnarray}
b^{r}(\tau)&=&\frac{-R(n_0^{\theta})^2}{2f_2^{-1}}\cos(2\omega_{\theta}\tau)\nonumber\\
&+&\frac{(n_0^{r})^2}{2Rf_6}\cos(2\omega\tau)\bigg\lbrace\left[\left(2+\frac{5M}{R}-\frac{6Q^2}{R^2}\right)-2\left(\frac{Q^2}{R^2}-\frac{Q^4}{R^4}\right)\left(\frac{M}{R}-\frac{Q^2}{R^2}\right)^{-1}\right]\nonumber\\
&-&\frac{a}{R}\sqrt{\frac{M}{R}-\frac{Q^2}{R^2}}\left[12-\frac{4Q^2}{R^2}\left(\frac{M}{R}-\frac{Q^2}{R^2}\right)^{-1}\right]+\frac{a^2}{R^2}\left[5-\frac{3Q^2}{R^2}\left(\frac{M}{R}-\frac{Q^2}{R^2}\right)^{-1}\right]\bigg\rbrace
,\label{42}\end{eqnarray}
\begin{eqnarray}
b^{\theta}(\tau)=\frac{(n_{0}^{r})(n_{0}^{\theta})}{r}\left[\left(2\sqrt{\frac{f_{3}}{f_{6}}}+1\right)\cos[(\omega_{\theta}-\omega)\tau]
-\left(2\sqrt{\frac{f_{3}}{f_{6}}}-1\right)\cos[(\omega_{\theta}+\omega)\tau]\right],
\label{43}
\end{eqnarray}
\begin{eqnarray}
b^{\varphi}(\tau)&=&\frac{(n_{0}^{\theta})^{2}\sin(2\omega_{\theta} \tau)}{\sqrt{f_{3}}}\left[1-\frac{a}{R}\sqrt{\frac{M}{R}-\frac{Q^{2}}{R^{2}}}\left(2+\frac{Q^{2}}{R^{2}}\left(\frac{M}{R}-\frac{Q^{2}}{R^{2}}\right)^{-1}\right)\right]\nonumber\\
&+&\frac{(n_{0}^{r})^{2}\sin(2\omega \tau)}{2R^{2}f_{2}f_{6}^{\frac{3}{2}}}\bigg\lbrace\left[\left(1-\frac{8M}{R}\right)\left(1-\frac{2M}{R}\right)+\frac{6Q^{2}}{R^{2}}-\frac{22MQ^{2}}{R^{3}}+
\frac{9Q^{4}}{R^{4}}\right.\nonumber\\
&+&\left.\left(\frac{Q^{2}}{R^{2}}-\frac{2Q^{4}}{R^{4}}+\frac{Q^{6}}{R^{6}}\right)\left(\frac{M}{R}-\frac{Q^{2}}{R^{2}}\right)^{-1}\right]+\frac{2a}{R}\sqrt{\frac{M}{R}-\frac{Q^{2}}{R^{2}}}\left[\left(8-\frac{25M}{R}+\frac{21Q^{2}}{R^{2}}\right)\right.\nonumber\\
&-&\left.\frac{1}{2}\left(\frac{Q^{2}}{R^{2}}-\frac{3Q^{4}}{R^{4}}\right)\left(\frac{M}{R}-\frac{Q^{2}}{R^{2}}\right)^{-1}-\frac{1}{2}\left(\frac{Q^{4}}{R^{4}}-\frac{Q^{6}}{R^{6}}\right)\left(\frac{M}{R}-\frac{Q^{2}}{R^{2}}\right)^{-2}\right]\nonumber\\
&-&\frac{a^{2}}{R^{2}}\left[\left(5-\frac{46M}{R}+\frac{59Q^{2}}{R^{2}}\right)-3\left(\frac{Q^{2}}{R^{2}} -\frac{Q^{4}}{R^{4}}\right)\left(\frac{M}{R}-\frac{Q^{2}}{R^{2}}\right)^{-1}\right]\nonumber\\
&-&\frac{a^{3}}{R^{3}}\sqrt{\frac{M}{R}-\frac{Q^{2}}{R^{2}}}\left[ 14-\frac{9Q^{2}}{R^{2}}\left(\frac{M}{R}-\frac{Q^{2}}{R^{2}}\right)^{-1}+\frac{Q^{4}}{R^{4}}\left(\frac{M}{R}-\frac{Q^{2}}{R^{2}}\right)^{-2}\right]\bigg\rbrace .
\label{44}
\end{eqnarray}
To obtain the geodesic curve $x^{\mu}$, we calculate $\frac{1}{2}\delta^{2}x^{\mu}$ with the result
\begin{eqnarray}
\delta^{2}t =(\delta^{2}t _{0})\tau+\delta^{2}t _{2r}\sin(2\omega \tau)+\delta^{2}t _{2\theta}\sin(2\omega_{\theta} \tau),\label{45}
\end{eqnarray}
\begin{eqnarray}
\delta^{2}r=\delta^{2}r_{0}+\delta^{2}_{2r}\cos(2\omega \tau),\label{46}
\end{eqnarray}
\begin{eqnarray}
\delta^{2}\theta =\delta^{2}\theta _{-}\cos[(\omega -\omega_{\theta})\tau]+\delta^{2}\theta _{+}\cos[(\omega +\omega_{\theta})\tau],\label{47}
\end{eqnarray}
\begin{eqnarray}
\delta^{2}\varphi =\delta^{2}\varphi _{0}\tau+\delta^{2}\varphi_{2r}\sin(2\omega \tau)+\delta^{2}\varphi_{2\theta}\sin(2\omega_{\theta} \tau).\label{48}
\end{eqnarray}
Coefficients in the above equations can be obtained after long calculations and are listed in the Appendix. As we expect in the Schwarzschild limit $a=0$, $Q=0$, and Kerr limit $Q=0$, the solution reduces to that of references \cite{Kerner} and \cite{Colistete}, respectively. By adding the second-order deviation vector $b^{\mu}(\tau)$ (with angular frequency $2\omega$) to the circular orbit we can get a better approximation of the elliptical orbit. The resulting trajectory is not an ellipse since $R\neq a$ and $e\neq\frac{n_0^{r}}{R}$, but for $x^{\mu}$ the perihelion and aphelion can be derived from $\omega\tau=2k\pi$ and $w\tau=(1+2k\pi)$ where $k\in Z$, respectively. Therefore the semi-major axis $a$ and eccentricity $e$ of an ellipse are given by
\begin{eqnarray}
a=R&-&\frac{(n_0^r)^2}{2Rf_6}\bigg\lbrace\left[\left(1-\frac{7M}{R}+\frac{5Q^{2}}{R^{2}}\right)+\left(\frac{Q^{2}}{R^{2}}-\frac{Q^{4}}{R^{4}}\right)\left(\frac{M}{R}-\frac{Q^{2}}{R^{2}}\right)^{-1}\right]\nonumber\\
&+&\frac{10a}{R}\sqrt{\frac{M}{R}-\frac{Q^{2}}{R^{2}}}\left[1-\frac{Q^{2}}{5R^{2}}\left(\frac{M}{R}-\frac{Q^{2}}{R^{2}}\right)^{-1}\right] -\frac{4a^{2}}{R^{2}} \left[1-\frac{Q^{2}}{2R^{2}}\left(\frac{M}{R}-\frac{Q^{2}}{R^{2}}\right)^{-1}\right]\bigg\rbrace ,\label{49}
\end{eqnarray}
\begin{eqnarray}
e&=&\frac{n_0^r}{R}-\frac{2(n_0^r)f_6}{\frac{(n_0^r)^2}{R}}\bigg\lbrace\left[\left(1-\frac{7M}{R}+\frac{5Q^{2}}{R^{2}}\right)+\left(\frac{Q^{2}}{R^{2}}-\frac{Q^{4}}{R^{4}}\right)\left(\frac{M}{R}-\frac{Q^{2}}{R^{2}}\right)^{-1}\right]\nonumber\\
&+&\frac{10a}{R}\sqrt{\frac{M}{R}-\frac{Q^{2}}{R^{2}}}\left[1-\frac{Q^{2}}{5R^{2}}\left(\frac{M}{R}-\frac{Q^{2}}{R^{2}}\right)^{-1}\right] -\frac{4a^{2}}{R^{2}} \left[1-\frac{Q^{2}}{2R^{2}}\left(\frac{M}{R}-\frac{Q^{2}}{R^{2}}\right)^{-1}\right]\bigg\rbrace^{-1}\nonumber\\
&=&\frac{n_0^r}{R}+{\cal O}\left(\frac{(n_0^r)^3}{R^3}\right).\label{50}
\end{eqnarray}
Finally, $\tau$ can be substituted in $r(\tau)$, resulting in $r(\varphi)$ up to the second order
\begin{eqnarray}
\frac{r(\varphi)}{R}=1&-&\left(\frac{n_0^r}{R}\right)\cos\left(\frac{\omega}{\omega _0}\varphi\right)+\left(\frac{n_0^r}{R}\right)^2\biggl[-1+\frac{1}{2f_6}\left[1-\frac{5M}{R}+\frac{Q^2}{R^2}+\left(\frac{Q^2}{R^2}-\frac{Q^4}{R^4}\right)\left(\frac{M}{R}-\frac{Q^2}{R^2}\right)^{-1}\right.\nonumber\\
&+&\left.\frac{6a}{R}\sqrt{\frac{M}{R}-\frac{Q^2}{R^2}}\left(1+\frac{Q^2}{3R^2}(\frac{M}{R}-\frac{Q^2}{R^2})^{-1}\right)+\frac{2a^2}{R^2}\right]\cos\left(\frac{2\omega}{\omega _0}\varphi\right)\biggr]+\cdots .\label{51}
\end{eqnarray}
In the Schwarzschild limit, we have an elliptical orbit
\begin{eqnarray}
{\frac{r(\varphi)}{R}}=1-\left(\frac{n_0^r}{R}\right)\cos\left(\frac{\omega}{\omega _0}\varphi\right)+\left(\frac{n_0^r}{R}\right)^2\left[-1+\frac{(1-\frac{5M}{R})}{2(1-\frac{6M}{R})}\cos\left(\frac{2\omega}{\omega _0}\varphi\right)\right]+\cdots, \label{Sh}
\end{eqnarray}
which is different from analogous result \cite{Kerner} in a constant term, because we have used another initial conditions. Then for the above result we obtain
\begin{figure}
\centering
\includegraphics[width=4.0in]{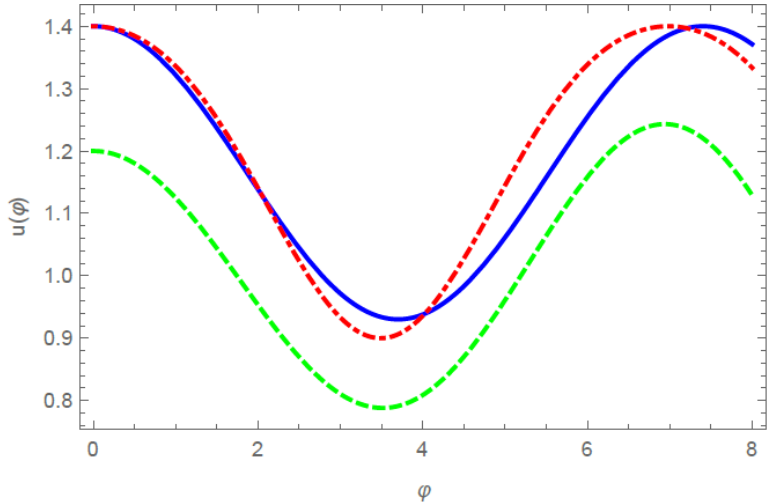}
\caption{The orbital motion $u(\varphi)$ for the numerical solution of the relativistic Binet's equation, $ u''(\varphi)+u(\varphi)=\frac{M}{h^2}+3Mu^2(\varphi) $, solid curve, its approximate solution, dashed curve, \cite{Inverno} and the orbital motion for the higher-order geodesic deviation method, equation (53), dashed-dotted curve. We set $M = 0.04$, $h = 0.2$, $\frac{\omega}{\omega_0} = 0.9$, $\frac{M}{R} = 0.032$, and $e = 0.2$.}
\label{orbital}
\end{figure}
\begin{eqnarray}
a&=&R-\frac{(n_0^r)^2\left(1-\frac{7M}{R}\right)}{2R\left(1-\frac{6M}{R}\right)},\label{53}
\end{eqnarray}
and
\begin{eqnarray}
e=\frac{2n_0^r\left(1-\frac{6M}{R}\right)}
{2R\left(1-\frac{6M}{R}\right)-\frac{(n_0^{r})^2}{R}\left(1-\frac{7M}{R}\right)}.\label{54}
\end{eqnarray}
Since the variable $ u(\varphi)=\frac{1}{r} $ is more convenient than $ r $, we compare the numerical solution of the relativistic version of Binet's equation and its approximate solution for $u(\varphi)$ with the $u(\varphi)$ obtained via the higher-order geodesic deviation method, equation (\ref{Sh}), in figure 3. As can be seen, our result has more coincidence with numerical graph.

Also for the Schwarzschild case the shape of the orbit in the limit $\frac{M}{R}\rightarrow0$ is described up to second-order in $e=\frac{n_0^r}{R}$ by
\begin{eqnarray}
r=\frac{r_0}{1+e\cos\varphi}=\frac{R(1-\frac {3}{2}e^2)}{1+e\cos\varphi}=R\left[1-e \cos\varphi +e^2 \left(-1+\frac{1}{2}\cos2\varphi\right)+\cdots\right],\label{55}
\end{eqnarray}
which is an exact equation of an ellipse with $a\simeq R(1-\frac{e^2}{2})$, in agreement with equations (\ref{49}) and (\ref{50}). In the third-order approximation, one may find a covariant deviation equation for $\delta^{3}x^{\mu}$ (see equation (72) of \cite{Kerner}), which can be written in matrix form. As previously mentioned, this matrix has the same left-hand side as those of equations (\ref{17}) and (\ref{36}), but a different vector in the right-hand side. Components of this vector contain terms including sine and cosine functions of the form $\omega\tau$ ($\omega_{\theta}\tau$) and $3\omega\tau$ ($3\omega_{\theta}\tau $). The presence of the functions $\cos(\omega\tau)$ and $\cos(\omega_{\theta}\tau)$, are the same as resonance terms, leading to solutions proportional to $\tau$. Since such terms in the solutions correspond to unbounded deviations, we can use Poincare's method to avoid this problem \cite{Kerner}.

\section{Conclusions}
In this paper we have generalized the geodesic deviation method \cite{Kerner} to the Kerr-Newman space-time which represents the most general black-hole. Our method is  fully relativistic and constitutes an alternative approach for calculating the perihelion precession and trajectory of test particles in gravitational fields. We started with a circular orbit with constant velocities. In this case the first-order geodesic deviations, equations (\ref{3}), was reduced to a system of linear differential equations with constant coefficients which is a collection of harmonic oscillators, equation (\ref{17}). The discrepancy between two circular frequencies, the unperturbed circular orbit $\omega _0$ and the perturbed orbit $\omega$ to first order, equations (\ref{16}) and (\ref{23}), is exactly what produces the precession of planet's perihelion. The effect of mass of a black-hole, the angular momentum and  electric charge of the source on the orbital precession was also discussed. We then proceeded to derive the second-order geodesic deviations, equation (\ref{9}), leading to a collection of harmonic oscillators excited by the external periodic forces represented by the right hand side of equation (\ref{36}), which may be solved easily. Finally, by adding the first corrections (linear approximation), second and higher-order corrections (non-linear approximation) to the circular orbit (zeroth-order approximation), we  constructed the trajectory of neutral test particles, equation (\ref{51}), in the Kerr-Newman space-time. Also, the study of non-equatorial orbits and the precession of the orbital plane in such a space time, using geodesic deviation method, would be an extension of our results for a future work.

Finally, it is worth stressing that in contrast to the Post-Newtonian expansion \cite{post1, post2, post3} where the orbits are calculated in an almost Minkowski space-time and the corrections due to the non-zero curvature are added perturbatively, there is no such constraints on the curvature of space-time in the geodesic deviation method and the fully relativistic effects are taken into account. Also, the additional terms of the form $M/R$ obtained in the geodesic deviation method when compared to similar solutions in the standard approximation result in  better accuracies. The other advantage of the geodesic deviation method is to provide a way for calculating small-eccentricity orbits in the vicinity of massive and compact objects. This is so because in the Post-Newtonian approximation, the assumption of weak gravitational fields and small velocities would  not be satisfied in scenarios  where the ratio $M/R$ is large.

\section*{Acknowledgements}
We would like to thank Richard Kerner for helpful discussions and Mohsen Khodadi for a careful reading of the manuscript and useful suggestions. We also thank the anonymous referee for valuable comments.

\section*{Appendix}
The coefficients of $\delta^2x^{\mu}$ in the second-order geodesic deviation equations are
\begin{eqnarray}
\delta^{2}t _{0}=0,\hspace{.5cm}\delta^{2}t _{2\theta}=\frac{(n_{0}^{\theta})^{2}}{\sqrt{f_{3}}}\frac{a^{2}}{R}\sqrt{\frac{M}{R}-\frac{Q^{2}}{R^{2}}}\left[ 1+\frac{Q^{2}}{2R^{2}}\left(\frac{M}{R}-\frac{Q^{2}}{R^{2}}\right)^{-1}\right],
\end{eqnarray}
\begin{eqnarray}
\delta^{2}t _{2r}&=&\frac{(n_{0}^{r})^{2}\sqrt{\frac{M}{R}-\frac{Q^{2}}{R^{2}}}}{Rf_{2}^2f_{6}^{\frac{3}{2}}}\bigg\lbrace\left[\left(2-\frac{15M}{R}+\frac{14M^{2}}{R^{2}}+\frac{10Q^{2}}{R^{2}}-\frac{17MQ^{2}}{R^{3}}+\frac{4Q^{4}}{R^{4}}\right)\right.\nonumber\\
 &+&\left.\frac{1}{2}\left(\frac{3Q^{2}}{R^{2}}-\frac{2Q^{4}}{R^{4}}-\frac{Q^{6}}{R^{6}}\right)\left(\frac{M}{R}-\frac{Q^{2}}{R^{2}}\right)^{-1}-\frac{1}{2}\left(\frac{Q^{4}}{R^{4}}-\frac{2Q^{6}}{R^{6}}+\frac{Q^{8}}{R^{8}}\right)\left(\frac{M}{R}-\frac{Q^{2}}{R^{2}}\right)^{-2}\right] \nonumber\\
&+&\frac{a}{R}\sqrt{\frac{M}{R}-\frac{Q^{2}}{R^{2}}}\left[\left(11+\frac{34M}{R}-\frac{64M^{2}}{R^{2}}+\frac{15Q^{2}}{R^{2}}+\frac{38MQ^{2}}{R^{3}}-\frac{26Q^{4}}{R^{4}}\right)\right.\nonumber\\
&-&\left.\frac{1}{2}\left(\frac{31Q^{2}}{R^{2}}-\frac{64Q^{4}}{R^{4}}+\frac{33Q^{6}}{R^{6}}\right)\left(\frac{M}{R}-\frac{Q^{2}}{R^{2}}\right)^{-1}-\frac{1}{2}\left(\frac{5Q^{4}}{R^{4}}-\frac{10Q^{6}}{R^{6}}+\frac{5Q^{8}}{R^{8}}\right)\left(\frac{M}{R}-\frac{Q^{2}}{R^{2}}\right)^{-2}\right]\nonumber\\
&-&\frac{a^{2}}{R^{2}}\left[\left(3+\frac{84M}{R}-\frac{126M^{2}}{R^{2}}-\frac{53Q^{2}}{R^{2}}+\frac{137MQ^{2}}{R^{3}}-\frac{36Q^{4}}{R^{4}}\right)\right.\nonumber\\
&-&\left.\frac{1}{2}\left(\frac{15Q^{2}}{R^{2}}-\frac{10Q^{4}}{R^{4}}-\frac{Q^{6}}{R^{6}}\right)\left(\frac{M}{R}-\frac{Q^{2}}{R^{2}}\right)^{-1}+\frac{1}{2}
\left(\frac{3Q^{4}}{R^{4}}-\frac{4Q^{6}}{R^{6}}+\frac{Q^{8}}{R^{8}}\right)\left(\frac{M}{R}-\frac{Q^{2}}{R^{2}}\right)^{-2}\right]\nonumber\\
&+&\frac{a^{3}}{R^{3}}\sqrt{\frac{M}{R}-\frac{Q^{2}}{R^{2}}}\left[\left(56-\frac{66M}{R}+\frac{35Q^{2}}{R^{2}}\right)-\left(\frac{16Q^{2}}{R^{2}}-\frac{15Q^{4}}{R^{4}}\right)\left(\frac{M}{R}-\frac{Q^{2}}{R^{2}}\right)^{-1}\right.\nonumber\\
&-4&\left.\left(\frac{Q^{4}}{R^{4}}-\frac{Q^{6}}{R^{6}}\right)\left(\frac{M}{R}-\frac{Q^{2}}{R^{2}}\right)^{-2}\right]-\frac{a^{4}}{R^{4}}\left[\left(12+\frac{21M}{R}-\frac{19Q^{2}}{R^{2}}\right)\right.\nonumber\\
&-&\left.\frac{1}{2}\left(\frac{21Q^{2}}{R^{2}}-\frac{8Q^{4}}{R^{4}}\right)
\left(\frac{M}{R}-\frac{Q^{2}}{R^{2}}\right)^{-1}+\frac{1}{2}\left(\frac{3Q^{4}}{R^{4}}-\frac{2Q^{6}}{R^{6}}\right)\left(\frac{M}{R}-\frac{Q^{2}}{R^{2}}\right)^{-2}\right]\nonumber\\
&+&\frac{a^5}{R^5}\sqrt{\frac{M}{R}-\frac{Q^2}{R^2}}\left[29-\frac{17Q^2}{2R^2}\left(\frac{M}{R}-\frac{Q^2}{R^2}\right)^{-1}-\frac{3Q^4}{2R^4}\left(\frac{M}{R}-\frac{Q^2}{R^2}\right)^{-2}\right]\nonumber\\
&-&\frac{a^6}{R^6}\left[27-\frac{9Q^2}{2R^2}\left(\frac{M}{R}-\frac{Q^2}{R^2}\right)^{-1}+\frac{Q^4}{2R^4}\left(\frac{M}{R}-\frac{Q^2}{R^2}\right)^{-2}\right]\bigg\rbrace
\end{eqnarray}
\begin{eqnarray}
\delta^{2}r_{0}=0,\hspace{.5cm}\delta^{2}r_{2r}&=&\frac{-(n_{0}^{r})^{2}}{Rf_{6}}\bigg\lbrace\left[\left(1-\frac{7M}{R}+\frac{5Q^{2}}{R^{2}}\right)+\left(\frac{Q^{2}}{R^{2}}-\frac{Q^{4}}{R^{4}}\right)\left(\frac{M}{R}-\frac{Q^{2}}{R^{2}}\right)^{-1}\right]\nonumber\\
&+&\frac{10a}{R}\sqrt{\frac{M}{R}-\frac{Q^{2}}{R^{2}}}\left[1-\frac{Q^{2}}{5R^{2}}\left(\frac{M}{R}-\frac{Q^{2}}{R^{2}}\right)^{-1}\right]\nonumber\\
&-&\frac{4a^{2}}{R^{2}}\left[1-\frac{Q^{2}}{2R^{2}}\left(\frac{M}{R}-\frac{Q^{2}}{R^{2}}\right)^{-1}\right]\bigg\rbrace ,
\end{eqnarray}
\begin{eqnarray}
\delta^{2}\theta _{-}=-\delta^{2}\theta _{+}=\frac{2(n_{0}^{r})(n_{0}^{\theta})}{R}\frac{\sqrt{f_3}}{\sqrt{f_6}}
\end{eqnarray}
\begin{eqnarray}
\delta^{2}\varphi _{0}=0,\hspace{.5cm}\delta^{2}\varphi _{2\theta}=\frac{(n_{0}^{\theta})^{2}}{\sqrt{f_{3}}}\left[ 1-\frac{2a}{R}\sqrt{\frac{M}{R}-\frac{Q^{2}}{R^{2}}} \left(1+\frac{Q^{2}}{2R^{2}} \left(\frac{M}{R}-\frac{Q^{2}}{R^{2}}\right)^{-1}\right)\right],
\end{eqnarray}
\begin{eqnarray}
\delta^{2}\varphi_{2r}&=&\frac{-2(n_{0}^{r})^{2}}{R^{2}f^{2}_{2}f_{6}^{\frac{3}{2}}}\bigg\lbrace\left[\left(5-\frac{32M}{R}\right)\left(1-\frac{2M}{R}\right)^{2}-\frac{9Q^{2}}{R^{2}}+\frac{140MQ^{2}}{R^{3}}-\frac{204M^{2}Q^{2}}{R^{4}}\right.\nonumber\\
&-&\left.\frac{63Q^{4}}{R^{4}}+\frac{128MQ^{4}}{R^{5}}-\frac{19Q^{6}}{R^{6}} -5\left(\frac{Q^{2}}{R^{2}}-\frac{3Q^{4}}{R^{4}}+\frac{3Q^{6}}{R^{6}}-\frac{Q^{8}}{R^{8}}\right)\left(\frac{M}{R}-\frac{Q^{2}}{R^{2}}\right)^{-1}\right]\nonumber\\
&-&\frac{2a}{R}\sqrt{\frac{M}{R}-\frac{Q^{2}}{R^{2}}}\left[ \left(26-\frac{119M}{R}+\frac{126M^{2}}{R^{2}}+\frac{66Q^{2}}{R^{2}}-\frac{137MQ^{2}}{R^{3}}+\frac{36Q^{4}}{R^{4}}\right)\right.\nonumber\\
&+&\left.\frac{1}{2}\left(\frac{3Q^{2}}{R^{2}}-\frac{2Q^{4}}{R^{4}}-\frac{Q^{6}}{R^{6}}\right)\left(\frac{M}{R}-\frac{Q^{2}}{R^{2}}\right)^{-1} -\frac{1}{2}\left(\frac{Q^{4}}{R^{4}}-\frac{2Q^{6}}{R^{6}}+\frac{Q^{8}}{R^{8}}\right)\left(\frac{M}{R}-\frac{Q^{2}}{R^{2}}\right)^{-2}\right]\nonumber\\
&+&\frac{2a^{2}}{R^{2}}\left[ \left(8-\frac{61M}{R}+\frac{66M^{2}}{R^{2}}-\frac{101MQ^{2}}{R^{3}}+\frac{20Q^{4}}{R^{4}}+\frac{68Q^{2}}{R^{2}}\right)-4\left(\frac{Q^{2}}{R^{2}}-\frac{2Q^{4}}{R^{4}}+\frac{Q^{6}}{R^{6}}\right)\right.\nonumber\\
&&\left.\left(\frac{M}{R}-\frac{Q^{2}}{R^{2}}\right)^{-1}\right]+\frac{2a^{3}}{R^{3}}\sqrt{\frac{M}{R}-\frac{Q^{2}}{R^{2}}}\left[ \left(5+\frac{21M}{R}-\frac{19Q^{2}}{R^{2}}\right)-\left(\frac{6Q^{2}}{R^{2}}-\frac{4Q^{4}}{R^{4}}\right)\right.\nonumber\\
&&\left.\left(\frac{M}{R}-\frac{Q^{2}}{R^{2}}\right)^{-1}+\left(\frac{Q^{4}}{R^{4}}-\frac{Q^{6}}{R^{6}}\right)
\left (\frac{M}{R}-\frac{Q^{2}}{R^{2}}\right)^{-2}\right]+\frac{a^{4}}{R^{4}}\left[\left(5-\frac{58M}{R}+\frac{75Q^{2}}{R^{2}}\right)\right.\nonumber\\
&-&\left.\left(\frac{Q^2}{R^2}-\frac{Q^4}{R^4}\right)\left(\frac{M}{R}-\frac{Q^2}{R^2}\right)^{-1}\right]+\frac{a^{5}}{R^{5}}\sqrt{\frac{M}{R}-\frac{Q^{2}}{R^{2}}}\left[ 14-\frac{9Q^{2}}{R^{2}}\left(\frac{M}{R}-\frac{Q^{2}}{R^{2}}\right)^{-1}\right.\nonumber\\
&+&\left.\frac{Q^{4}}{R^{4}}\left(\frac{M}{R}-\frac{Q^{2}}{R^{2}}\right)^{-2}\right]\bigg\rbrace .
\end{eqnarray}

\end{document}